# Porosity Dependence of Powder Compaction Constitutive Parameters: Determination Based on Spark Plasma Sintering Tests

Charles Manière[a] and Eugene A. Olevsky[a, b]*

(a) Powder Technology Laboratory, San Diego State University, San Diego, USA
(b) NanoEngineering, University of California, San Diego, La Jolla, USA

**Keywords**
Sintering; visco-plastic model; bulk modulus; shear modulus; sinter-forging; spark plasma sintering

**Abstract**

   The modeling of powder compaction process, such as spark plasma sintering (SPS), requires the determination of the visco-plastic deformation behavior of the particle material including the viscosity moduli. The establishment of these parameters usually entails a long and difficult experimental campaign which in particular involves several hot isostatic pressing tests. A more straightforward method based on the coupled sinter-forging and die compaction tests, which can be easily carried out in a regular SPS device, is presented. Compared to classical creep mechanism studies, this comprehensive experimental approach can reveal the in situ porous structure morphology influence on the sintering process.

―――――――――――――――――――――
*Corresponding author: **EO**: Powder Technology Laboratory, San Diego State University, 5500 Campanile Drive, San Diego, CA 92182-1323,
Ph.: (619)-594-6329; Fax: (619)-594-3599, *E-mail address*: eolevsky@mail.sdsu.edu

The simulation of the powder compaction in advanced sintering techniques, such as spark plasma sintering, is a helpful research tool enabling the prediction and optimization of: the densification nonuniformity [1], the tooling resistance to the stress solicitation [2], and the elaboration of complex shapes [3]. For the SPS technology, these mechanical simulations are often coupled with the Joule heating modeling [4–10] in a multiphysics approach rendering a comprehensive prediction of this process' electro-thermal-mechanical phenomena [11–13]. As described by numerous authors [5,14–17], the main challenge related to the Joule heating modeling is to identify the non-ideal electric and thermal contacts in the SPS tooling-specimen setup as the dominant parameters controlling the temperature field distribution. Concerning the aspects of the mechanical modeling, the great challenge is to identify all the model constitutive parameters. The powder compaction model for both pressure and pressureless sintering techniques can be described by the general continuum theory of sintering [18]. For pressure assisted techniques such as SPS this approach can be reduced to the description of a visco-plastic porous body (a continuum made of a dense phase and porosity) behavior [3,19,20]. The dense phase nonlinear viscous behavior is often modeled via a power law creep. The stress/strain behavior of the porous medium is also described by the porosity-dependent shear and bulk moduli. The experimental determination of these moduli is usually rather cumbersome, therefore, as a rule, the values of these parameters are approximated theoretically to the detriment of the overall modeling accuracy [18].

Numerous theoretical derivations of the shear and bulk moduli consider linear viscous [21,22] or power law creep [23–25] materials with an idealized porous body "skeleton". As reported by Wolff *et al* [26] these theoretical moduli are often characterized by a good functional trend, but the difference of a considerable magnitude is observed between the theoretical and experimentally determined moduli values [19,27–29]. The consequence of this discrepancy is then a possible theoretical overestimation of the equivalent creep parameters. Failure in the identification of the creep mechanism has been reported [30] when using traditional isothermal linear regression methods [3,31–33]. The isothermal regime is very sensible and the somewhat inaccurate estimation of the shear and bulk moduli can generate a significant error in the sintering mechanism evaluation [30]. A more precise experimental determination of the shear and bulk moduli is therefore of high interest for the sintering modeling. However, the traditional methods of the determination of these moduli are very time consuming and require, inter alia, several instrumented hot isostatic pressing (HIP) tests [19,29], a rather expensive instrumentation to setup. An alternative solution using coupled sinter-forging and die compaction tests is described in this paper. This combination, inspired from refs. [27,28], can be easily adapted to a simple SPS machine such as for the study Ref [34,35] on Ti-6Al-4V and TiAl respectively. In these work, the creep and densification moduli are determined by in situ SPS approach. It is to be noted for small grain size ceramic powders, the in situ creep tests [34–40] are more suited as the SPS approach is able to more or less preserve fine microstructures [41–45].

To apply this method, the continuum theory of sintering needs to be reduced to its general analytical form for both the sinter-forging and die compaction tests. Considering the minimum operational pressure of 40 MPa and the 5 μm average particle size, the sintering stress can be neglected and the general formulation of the continuum theory of sintering gives the stress tensor $\underline{\sigma}$ the following expression [18]:

$$\underline{\sigma} = \frac{\sigma_{eq}}{\dot{\varepsilon}_{eq}}\left(\varphi\underline{\dot{\varepsilon}} + \left(\psi - \frac{1}{3}\varphi\right)\dot{e}\underline{\mathbb{1}}\right) \qquad (1)$$

where $\underline{\dot{\varepsilon}}$ is the strain rate tensor, $\underline{\mathbb{1}}$ is the identity tensor, $\varphi$ and $\psi$ the shear and bulk moduli to be determined, $\dot{\varepsilon}_{eq}$ and $\sigma_{eq}$ the equivalent strain rate and stress defined by [46]:

$$\dot{\varepsilon}_{eq} = \frac{1}{\sqrt{1-\theta}}\sqrt{\varphi\dot{\gamma}^2 + \psi\dot{e}^2} \qquad (2)$$

$$\sigma_{eq} = \frac{\sqrt{\frac{\tau^2}{\varphi} + \frac{P^2}{\psi}}}{\sqrt{1-\theta}} \qquad (3)$$

with $\theta$ being the porosity and the strain rate and stress tensor invariants given by:

$$\dot{\gamma} = \sqrt{2\left(\dot{\varepsilon}_{xy}^2 + \dot{\varepsilon}_{xz}^2 + \dot{\varepsilon}_{yz}^2\right) + \frac{2}{3}\left(\dot{\varepsilon}_x^2 + \dot{\varepsilon}_y^2 + \dot{\varepsilon}_z^2\right) - \frac{2}{3}\left(\dot{\varepsilon}_x\dot{\varepsilon}_y + \dot{\varepsilon}_x\dot{\varepsilon}_z + \dot{\varepsilon}_y\dot{\varepsilon}_z\right)} \qquad (4)$$

$$\tau = \sqrt{(\sigma_x - \sigma_y)^2 + (\sigma_y - \sigma_z)^2 + (\sigma_z - \sigma_x)^2 + 6(\sigma_{xy}^2 + \sigma_{yz}^2 + \sigma_{xz}^2)}/\sqrt{3} \qquad (5)$$

$$\dot{e} = \dot{\varepsilon}_x + \dot{\varepsilon}_y + \dot{\varepsilon}_z \text{ and } P = (\sigma_x + \sigma_y + \sigma_z)/3 = I_1/3 \qquad (6).$$

The porosity is determined locally by the mass conservation equation:

$$\frac{\dot{\theta}}{1-\theta} = \dot{\varepsilon}_x + \dot{\varepsilon}_y + \dot{\varepsilon}_z \qquad (7).$$

The equivalent strain rate and stress of the dense phase are related to each other via a creep power law:

$$\dot{\varepsilon}_{eq} = A\sigma_{eq}^n = A_0 exp\left(\frac{-Q}{RT}\right)\sigma_{eq}^n \qquad (8)$$

where for pure nickel [47,48]: $A_0 = 2.06E - 8 \; MPa^{-n}s^{-1}$, $Q = 171.1$ kJ mol$^{-1}$ and $n=7$.

The die compaction case (such as in traditional SPS configuration) is characterized by a unique displacement along z-axis (assuming compaction direction along z-axis) which gives the external strain rate tensor the following analytical approximation:

$$\underline{\dot{\varepsilon}} \equiv \begin{pmatrix} 0 & 0 & 0 \\ 0 & 0 & 0 \\ 0 & 0 & \dot{\varepsilon}_z \end{pmatrix} \qquad (9).$$

Replacing (9) in (2,4,6), one obtains the simplifications:

$$\dot{e} = \dot{\varepsilon}_z \; ; \; \dot{\gamma} = |\dot{\varepsilon}_z|\sqrt{\frac{2}{3}} \; ; \; W = |\dot{\varepsilon}_z|\sqrt{\frac{\psi + \frac{2}{3}\varphi}{1-\theta}} \qquad (10)$$

Which, using (1) and (8), renders the general analytical form of the die compaction loading mode:

$$|\dot{\varepsilon}_z| = A\left(\psi + \frac{2}{3}\varphi\right)^{\frac{-n-1}{2}}(1-\theta)^{\frac{1-n}{2}}|\sigma_z|^n \qquad (11).$$

The sinter-forging case is characterized by a unique loading along z-axis (assuming loading direction along z-axis) which gives the external stress tensor the following analytical approximation:

$$\underline{\sigma} \equiv \begin{pmatrix} 0 & 0 & 0 \\ 0 & 0 & 0 \\ 0 & 0 & \sigma_z \end{pmatrix} \qquad (12).$$

To determine the sinter-forging constitutive equation we need to determine first the strain rate tensor expression that depends on the stress tensor components.

Starting from (1) and considering the relationship $P\dot{\varepsilon}_{eq} = \sigma_{eq}\psi tr(\underline{\dot{\varepsilon}})$ for the stress and strain rate tensor invariants [46] one can determine:

$$\underline{\dot{\varepsilon}} = \frac{\dot{\varepsilon}_{eq}}{\sigma_{eq}}\left(\frac{\underline{\sigma}}{\varphi} - \left(\frac{1}{3\varphi} + \frac{1}{9\psi}\right)I_1\underline{\mathbb{1}}\right) \qquad (13).$$

Then, considering the stress tensor deviator expression $\underline{s} = \underline{\sigma} - I_1\underline{\mathbb{1}}/3$, we finally obtain the general form:

$$\underline{\dot{\varepsilon}} = \frac{\dot{\varepsilon}_{eq}}{\sigma_{eq}}\left(\frac{\underline{s}}{\varphi} + \frac{I_1}{9\psi}\underline{\mathbb{1}}\right) \qquad (14).$$

If we consider the simplification of (12) in (3,5,6) we obtain for sinter-forging:

$$\tau = \sqrt{\frac{2}{3}}|\sigma_z| \; ; \; P = -\frac{|\sigma_z|}{3} \; ; \; \sigma_{eq} = |\sigma_z|\frac{\sqrt{\frac{2}{3\varphi}+\frac{1}{9\psi}}}{\sqrt{1-\theta}} \; ; \; s_z = -\frac{2}{3}|\sigma_z| \qquad (15).$$

Using (14), (8) and (15), the final constitutive equation for sinter-forging is then:

$$|\dot{\varepsilon}_z| = A(1-\theta)^{\frac{1-n}{2}}\left(\frac{2}{3\varphi} + \frac{1}{9\psi}\right)^{\frac{n+1}{2}}|\sigma_z|^n \qquad (16).$$

Combining (11) and (16) it is possible to experimentally determine parameters $\varphi$ and $\psi$ at fixed porosity and temperature values. This can be achieved by solving the system of the two equations below where the first member is unknown ($\varphi$ and $\psi$), and the second member can be accessed experimentally by sinter-forging and die compaction tests ($\theta, |\dot{\varepsilon}_z|, |\sigma_z|$ are experimentally determined; $A, n$ are known by creep tests).

$$\begin{cases} \frac{2}{3\varphi} + \frac{1}{9\psi} = \left(|\dot{\varepsilon}_z|A^{-1}(1-\theta)^{\frac{n-1}{2}}|\sigma_z|^{-n}\right)^{\frac{2}{n+1}} & \text{sinter} - \text{forging} \\ \psi + \frac{2}{3}\varphi = \left(|\dot{\varepsilon}_z|^{-\frac{1}{n}}(1-\theta)^{\frac{1-n}{2n}}A^{\frac{1}{n}}|\sigma_z|\right)^{\frac{2n}{n+1}} & \text{Die compaction} \end{cases} \qquad (17)$$

Considering the sinter-forging case, it is obvious that the behavior of a loose powder specimen at a constant applied stress $|\sigma_z|$ can provoke the specimen's collapse and, in turn, a very high strain rate $|\dot{\varepsilon}_z|$. Consequently, in equation (16), the summation $\frac{2}{3\varphi} + \frac{1}{9\psi}$ tends to infinity, and both $\varphi$ and $\psi$ tend to zero at a critical porosity $\theta_c$ close to the porosity of a loose powder. Another fact is that the equivalent stress (3) tends to the von Mises stress expression at full specimen's density when $\varphi$ and $\psi$ tend to 2/3 and ∞, respectively. Similarly, the sinter-forging equation (16) is reduced to the forging equation (8) at full specimen's density and for $\varphi$ and $\psi$ equal to 2/3 and $\infty$, respectively.

All the experiments have been performed using an SPS machine (SPSS DR.SINTER Fuji Electronics model 515). The identification of $\varphi$ and $\psi$ has been conducted under isotherm isobar conditions. The nickel powder (Cerac, Ni 99.9% pure, 5 µm) was electrically insulated from the SPS tooling by a boron nitride spray to prevent any electric current effect disturbance. In this way, the die compaction tests are close to hot pressing conditions. The temperature was measured by a K type thermocouple directly at the specimen's surface for the sinter-forging tests. A separate experiment with a second thermocouple in the powder was carried out for the die compaction tests. For the sinter-forging tests pre-consolidated 10 mm diameter 5 mm height specimens with various relative densities were employed for each test. The sinter-forging tests were very short in time in order to measure a value of $|\sigma_z|$ and $|\dot{\varepsilon}_z|$ with a change of the relative density limited to about 1%; see the typical force and displacement patterns and the experimental setup configuration in figure 1a. The steps of the curve figure 1a reveal the overall configuration have a displacement detection accuracy of about 10 µm. For the die compaction tests, serval successive isobaric tests have been performed to cover a wide range of the relative density; see typical force and relative density patterns and the experimental setup configuration in figure 1b (where the compaction displacement can be directly convert into relative density with the instant and final specimen height hi, hf and relative density RDi, RDf relation: RDi=hf*RDf/hi).

Several sinter-forging tests have been carried out for the specimens with different initial relative densities; these tests rendered (using Eq. (17)) the points in figure 2 (upper left). Several die compaction tests have been performed at different constant temperatures; these tests rendered (using Eq. (17)) the points in figure 2 (upper right). These latter points converge to the same curve confirming for this powder the unicity of $\varphi$ and $\psi$ functions. Using the fitting of the points of the two graphs it is then possible to obtain $\varphi$ and $\psi$ by solving the set of the two equations (17), see figure 2 (lower). As expected, $\varphi$ and $\psi$ functions tend to 2/3 and ∞ at full density and the finalized fit of these functions is given below assuming the critical porosity of the powder to be $\theta_c = 0.55$ and their analytical expression inspired from ref. [19].

$$\varphi = \frac{1}{\left(\frac{3}{2}+5\left(\frac{\theta}{0.55-\theta}\right)^{1.2}\right)} \tag{18}$$

$$\psi = 0.36\left(\frac{0.55-\theta}{\theta}\right)^{0.6} \tag{19}$$

At this stage, the model is completely determined and different experiments in an-isotherm/isobar and isotherm/isobar conditions can be conducted to test the validity of the model determined above. All these experiments are different from the series of the experiments used to identify the model. The model predicts well the data points for the ramping experiments at 50 and 100 K/min shown in figure 3a with still some discrepancies at the end of sintering for the 50 K/min experiment. The isotherm regime (figures 3b, 3c, 3d) manifest a very good prediction for 550°C and some discrepancy for 660°C. Considering the high sensibility of the isotherm regime densification and the independency of the experiments, the error of the model relative density predictions (<5%) is acceptable for all the considered cases. In a previous study using theoretical moduli [3], we showed that the equivalent creep parameters may change depending on the pressure and heating rate. The present work shows that the determination of the experimental shear and bulk moduli from the fully dense material creep data provides a more generalized constitutive behavior description less sensible to the temperature and pressure regimes. The verification of the model is also extended to the forging of samples with high residual porosities. The residual porosity present in specimens or generated during the forging process can lead to a more ductile (deformable) behavior. This increase of the specimen ductility can be explained by the identified model and the sinter-forging equation (16). As shown in figure 4, the model/experiment comparison for the forged samples at different levels of the average residual porosity and temperature indicates a good agreement.

To conclude, the approach presented in this paper allows the identification of the shear and bulk moduli of porous materials using simple SPS-based experimental tests. The determined moduli seem to be only porosity dependent, and, compared to classical approaches, the resulting model is more stable when tested under very different regimes and experimental setup configurations. This approach should enable also experimental studies on the influence of the porous structure morphology evolution during SPS experiments (to be considered in future investigations), an aspect that is traditionally limited by theoretical approximations only.

**Acknowledgements**

The support of the US Department of Energy, Materials Sciences Division, under Award No. DE-SC0008581 is gratefully acknowledged.

Fig. 1: Displacement, relative density and loading profiles for (a) sinter-forging and (b) die compaction tests.

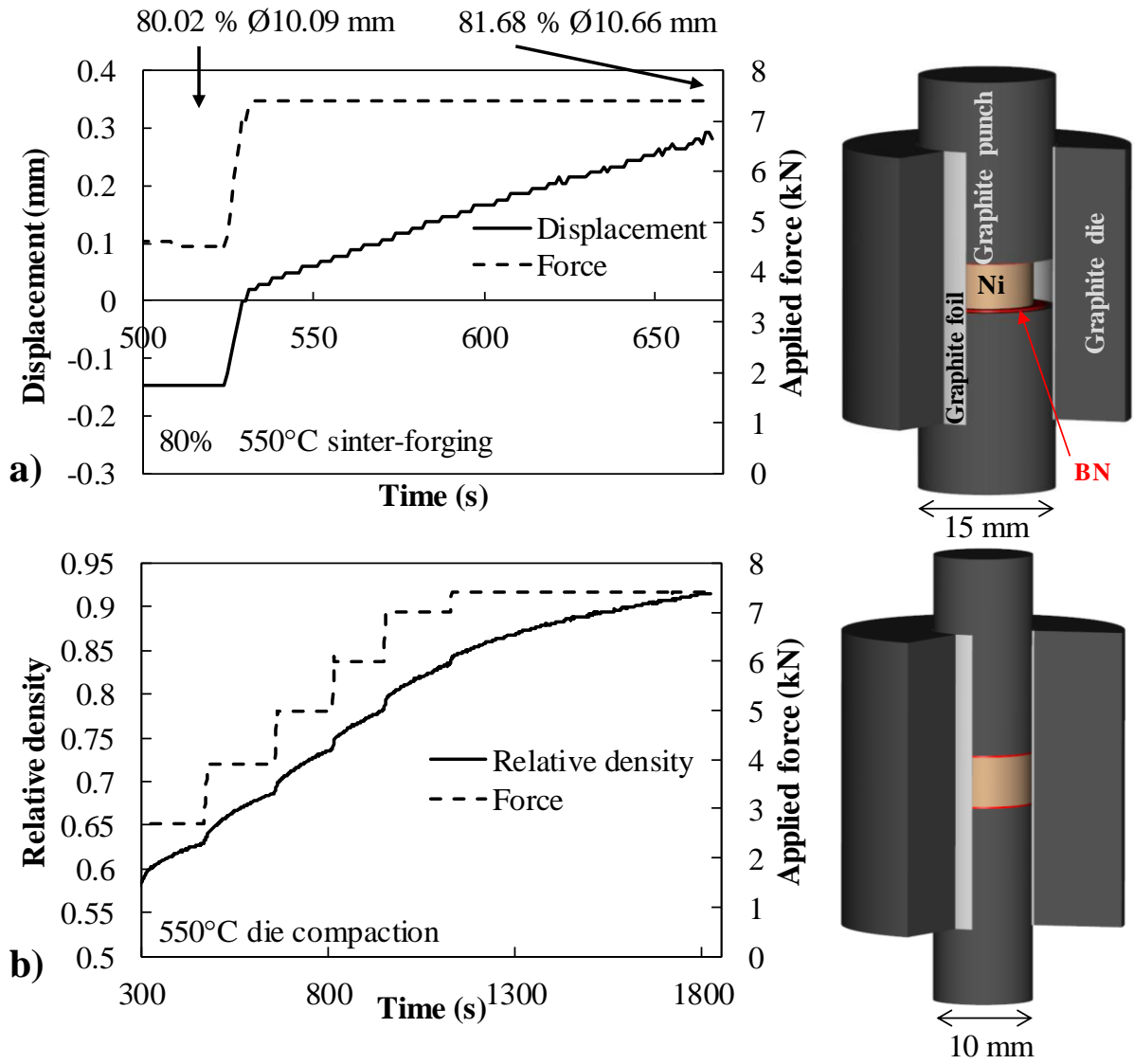

Fig.2: Porosity dependence of the constitutive parameters for sinter-forging (upper left), die compaction (upper right) tests and the resulting shear and bulk moduli (lower).

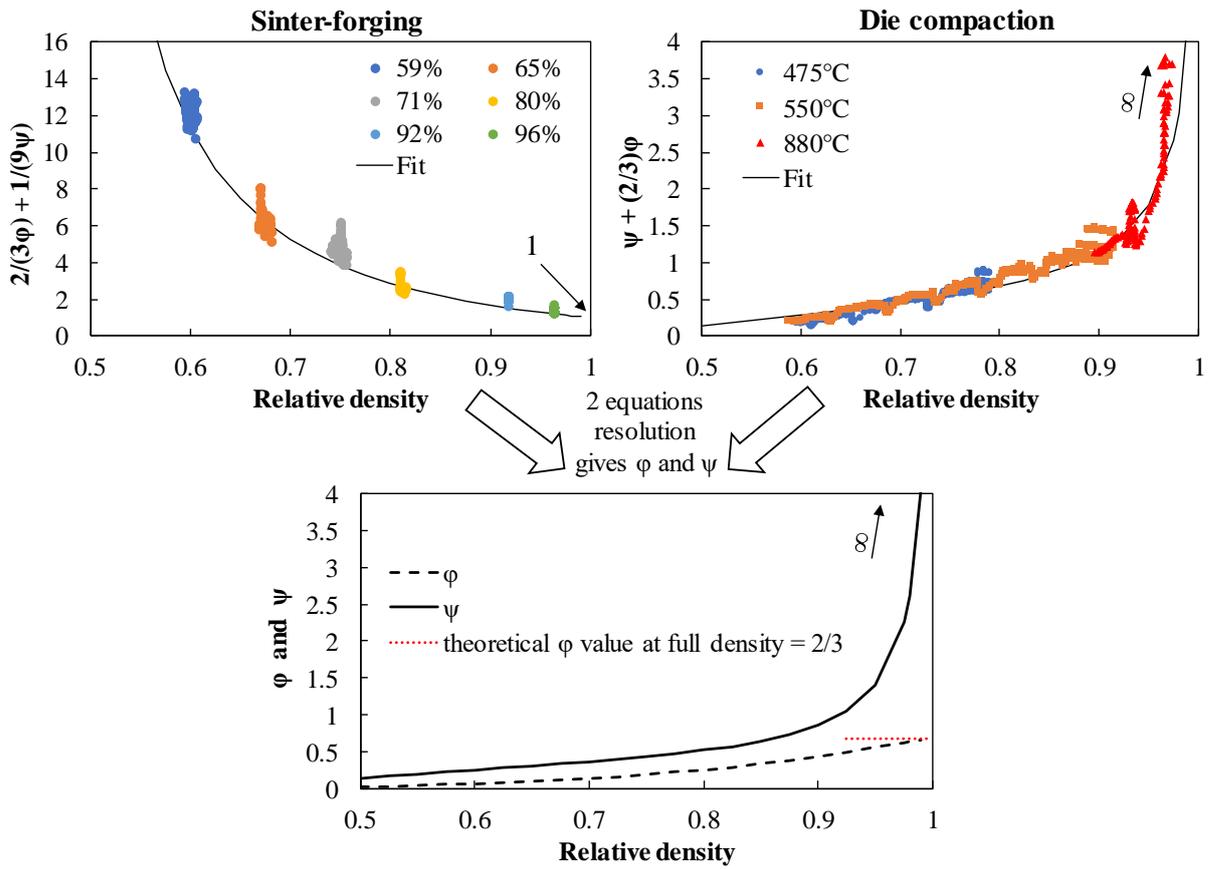

Fig.3: Model validation using independent die compaction tests in ramping (a) and isotherm temperature profiles 450°C (b), 550°C (c), 660°C (d).

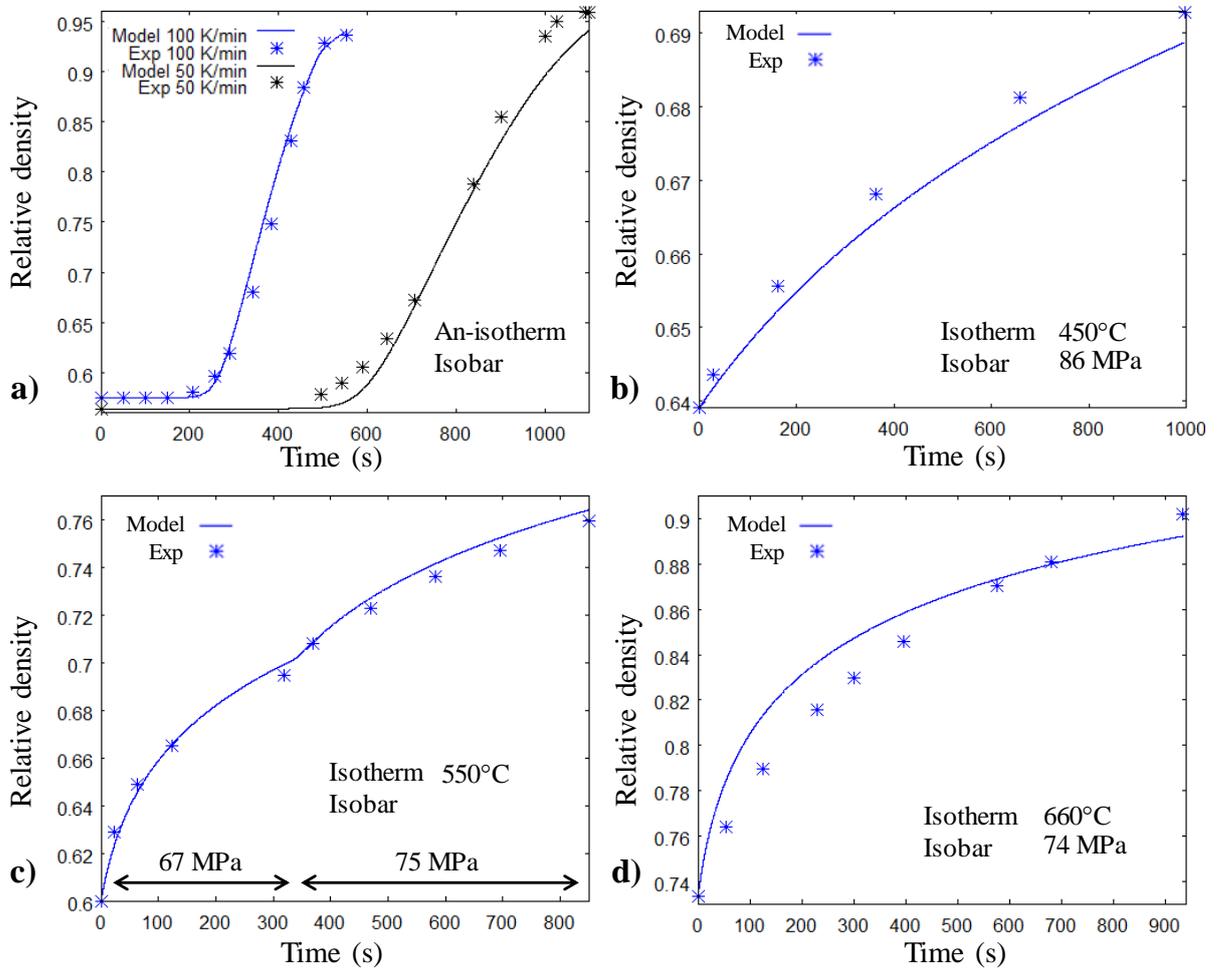

Fig.4: Model/experiment verification of ductility enhancement induced by high residual porosity level in forging experimental configuration at different temperatures and pressures (for each temperature the average specimen densification is indicated).

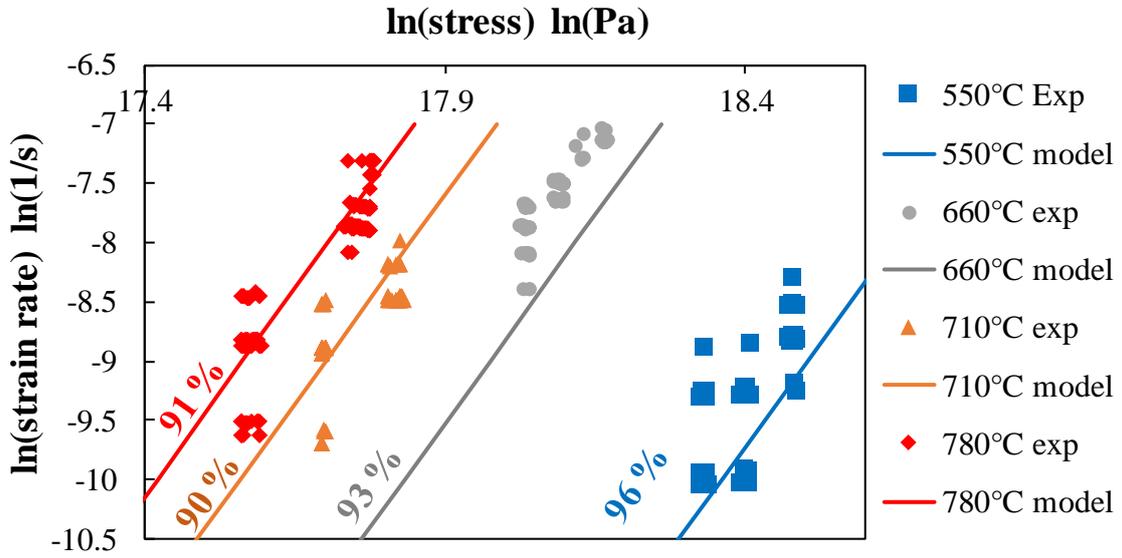